\documentstyle[multicol,aps,prl,floats]{revtex}
\draft
\begin{document}
\title{Two complementary descriptions of intermittency}
\author{E. Balkovsky$^a$ and G. Falkovich$^{a,b}$}
\address{$^a$ Physics of Complex Systems, Weizmann Institute of Science,
Rehovot 76100, Israel\\
$^b$ Institute for Advanced Study, Princeton, NJ 08540 USA}
\date{\today}
\maketitle

\begin{abstract} 
We describe two complementary formalisms designed for the description of
probability density function (PDF) of the gradients of turbulent fields.
The first approach, we call it adiabatic, describes PDF at the values much less
than dispersion. The second, instanton, approach gives the tails of PDF at the
values of the gradient much larger than dispersion. Together, both approaches
give satisfactory description of gradient PDFs, as illustrated here by an
example of a passive scalar advected by a one-dimensional compressible random 
flow. 
\end{abstract}

\begin{multicols}{2}
Probably the most striking feature of developed turbulence is its intermittent
spatial and temporal behavior. The structures that arise in a random flow
manifest themselves as high peaks at random places and at random times. The
intervals between them are characterized by a low intensity and a large size.
Rare high peaks are responsible for PDF tails while the regions of low
intensity contribute PDF near zero. That physical picture prompts an
attempt to describe intermittency at the level of a single-point PDF by two
complementary approaches. The first approach was recently
introduced to describe rare strong
fluctuations as optimal fluctuations realizing probability 
extrema\cite{96FKLM,96GM,97BFKL,96Chert,97FL}.  Called an instanton
approach, this formalism is based upon a path-integral representation of
conditional probability with optimal fluctuations being saddle points in the
integral.  
A counterpart to the instanton approach is suggested here for the
description of gradient PDF at small values, the approach is just an adiabatics
when high-order spatial derivatives are consistently neglected.

The center anomaly and tails of the
gradient PDF are two sides of the same coin called intermittency which is the
main target in modern turbulence studies.
It this Letter, we demonstrate how both methods applied together give a
consistent description of the gradient PDF. Note that the intermediate part of
the PDF (which is beyond our approaches) where the matching of the asymptotics
occurs is not that interesting because it is nonuniversal i.e.  depends on the
particular form of the pumping correlation function. The central peak and the
tail are robust, their form provides the main information on the probability of
both main body of the events and strong fluctuations.

Let us show how such a description can be developed by using, probably, the 
simplest 
(yet nontrivial) turbulent problem of a passive scalar $\theta(x,t)$
advected by one-dimensional random flow $v(x,t)$ which
is smooth in space and white in time (this is a compressible version 
\cite{97VM}
of a well-known Kraichnan model \cite{74Kra-a}):
\begin{equation}
\label{model}
\partial_t\theta+v\nabla\theta=\kappa\Delta\theta+ \phi .
\end{equation}
Both the velocity $v$ and the source function $\phi$ are 
supposed homogeneous, Gaussian and $\delta$-correlated in time:
$\langle \phi(x,t)\,\phi(x',t')\rangle=\chi({|x-x'|})
\delta(t-t')$
where $\chi$ is some function that decays on a scale $L$, the value $\chi(0)=P$
is the flux of $\theta^2$.
The correlation function
of the velocity may be defined by two parameters, typical velocity $V$ and 
correlation length $L_v$:
\begin{equation}\langle v(x,t)v(x',t')\rangle
=\bigl[VL_v-VL_v^{-1}(x-x')^2\bigr]\delta(t-t')\,.\label{vel}\end{equation}
At studying a simultaneous statistics, the coordinate-independent
part drops out. We assume $L_v\gg L$.

Let us first implement a simple adiabatic approach  neglecting diffusion term.
Then for the single-point PDF ${\cal 
P}(\omega,t)=\langle\delta[\theta_x(x,t)-\omega]\rangle$
one obtains a closed Fokker-Planck equation
\begin{equation}
2{\partial {\cal P}\over \partial t}=
(D\omega^2+T){\partial^2{\cal P}\over\partial 
\omega^2}+4D\omega{\partial{\cal P}\over\partial 
\omega}+2D{\cal P}\ ,\label{FP1}
\end{equation}
where we denote $T=\chi''(0)$ and $D=VL_v^{-1}$ the variances of $\phi_x$ and 
$v_x$
respectively. That equation has an equilibrium steady solution
\begin{equation}
{\cal P}(\omega)\propto(T+D\omega^2)^{-1}\label{ad}
\end{equation} 
expected to be applicable for $\omega^2\ll P/\kappa$. Since $T\simeq P/L^2$ and
the Peclet number $Pe^2=DL^2/4\kappa$ is assumed to be large then (\ref{ad})
has a wide interval of validity. Note that $T/D$ is a square gradient produced
by the pumping during the typical stretching time $D^{-1}$. For $Pe\gg1$,
$T/D\ll P/\kappa$. Limiting solutions obtained at $\omega^2\gg T/D$ and at 
$\omega^2\ll T/D$ by a time-separation procedure \cite{97CKVa} coincide with
(\ref{ad}). 

Let us now describe
the tail of the probability density function ${\cal P}(\omega)$
at $\omega^2\gg P/\kappa$. It is clear from (\ref{vel}) 
that the correlation functions of the strain field $\sigma=v_x$
are $x$-independent that is $\sigma$ can be treated as a
random function of time $t$ only. To exploit that, it is convenient to pass
into the comoving reference frame that is to the frame moving with the
velocity of a Lagrangian particle of the fluid \cite{96FKLM,97BFKL}.
The Martin-Siggia-Rose action $I$ for the $n$-th order moment
of the gradient $\theta_x$ is \cite{97CKVa}:
\begin{eqnarray}&&
{\cal I}=\int dtdx\,p\partial_t\theta-\int E dt
-\frac{in}{2}
\log(\theta_x)^2
\label{i1}
\\&&
E=\int\! dx\,p(-\sigma 
x\partial_x\theta+\kappa\partial_x^2\theta)\nonumber\\&&-
\frac{i}{2}\int\! dx_1dx_2\,p_1\chi(x_{12}) p_2
-\frac{i}{4D}(\sigma-D)^2
\label{energ}
\end{eqnarray}

Assuming $n\gg1$ we shall calculate the moment in the saddle-point
approximation $\langle\theta_x^n\rangle=\exp({{\cal I}_{extr}})$, 
see \cite{96FKLM,96GM,97BFKL,96Chert,97FL} for the details. 
Here, ${\cal I}_{extr}$ is to be calculated on the
flow configuration (optimal fluctuation or instanton)
that 
has to satisfy extremum equations for the action:
\begin{eqnarray}&&
\partial_t\theta+\sigma x\partial_x\theta-\kappa\partial_x^2\theta=
-i\int\!dx' p(t,x')\chi(x-x')\label{e1}\\&&
\partial_t p+\sigma\partial_x(x 
p)+\kappa\partial_x^2p=-y\delta(t)\delta'(x)\label{e2}\\&&
\sigma-D=2iD\int\!dx\, px\partial_x\theta \label{e3}
\end{eqnarray}
Here $y=-in/\theta_x(0,0)$. For calculations, it is more
convenient to use this auxiliary parameter instead of $\theta_x(0,0)$.
The boundary conditions
are $\theta(x,-\infty)=0$ and $p(x,+0)=0$ \cite{96FKLM}. 
The solution of (\ref{e1},\ref{e2})
can be sought in the following form
\begin{eqnarray}&&
\theta\!=\!f\left(\tau(t),x\sqrt{w(t)}\right),\ p\!=\!\sqrt{w(t)}\,
g\left(\tau(t),x\sqrt{w(t)}\right)\label{th1}\\&&
\partial_t w=-2\sigma w,\quad \partial_t\tau=-w\label{ata},\quad
w(0)=1,\quad\tau(0)=0.
\end{eqnarray}
The functions $f$ and $g$ satisfy the following equations
\begin{eqnarray}&&
\partial_\tau g-\kappa\partial_\xi^2 g=y\delta(\tau)\delta'(\xi)\\&&
\partial_\tau f+\kappa\partial_\xi^2 
f=\frac{i}{w(\tau)}\int_{-\infty}^{\infty}\! d\xi'\,
g(\tau,\xi')\chi\left(\frac{(\xi-\xi')}{\sqrt{w(\tau)}}\right)
\end{eqnarray}
The solution can be found in the Fourier-representation
\begin{eqnarray}
g(\tau,k)=&&iky e^{-\kappa k^2\tau}
\label{pp}\\
f(\tau,k)=&&-ky\int\limits_{\tau_0}^{\tau}\frac{d\tau'}{\sqrt{w(\tau')}}
\chi\left(k\sqrt{w(\tau')}\right)e^{\kappa k^2(\tau-2\tau')}
\label{f}\label{tt}\\
\sigma-D=&&-2Dy^2\int_{\tau_0}^{\tau}\!\,\frac{d\tau'}{\sqrt{w(\tau')}}
\int_{-\infty}^{\infty}\!
\frac{dk}{2\pi}\, k^2(1-2\kappa k^2\tau)
\nonumber\\&&\times\chi\left(k\sqrt{w(\tau')}\right)
e^{-2\kappa k^2\tau'}\label{i2}
\end{eqnarray}
Here $\tau_0$ is the maximal value for $\tau$, which is determined by
the moment when the following integral diverges
\begin{eqnarray}&&
t(\tau)=-\int^{\tau}_0 \!\frac{d\tau}{w(\tau)}.
\label{t}\end{eqnarray}

In the following we will work in the dimensionless units. We put $D=1$,
$P\equiv\chi(0)=1$, and $L=1$. 
Then $\kappa=1/(4Pe^2)$, where $Pe$ is the Peclet number. We believe $Pe\gg 1$.
Calculating $\partial_x\theta(0,0)$ from (\ref{f}) 
we obtain the following self-consistency condition for $y$
\begin{eqnarray}&&
\frac{n}{|y|^2}=
\int_0^{\tau_0}
\frac{d\tau}{w^2(\tau)}
\phi\left(\frac{\tau}{Pe^2w(\tau)}\right)\,.
\label{nlc}\\&&
\phi(x)=\int_{-\infty}^\infty\frac{dk}{2\pi}k^2\chi(k)
\exp\left(-\frac{k^2x}{2}\right)
\end{eqnarray}
The function $\phi(x)$ has the following asymptotics:
$\phi(x)\to 1$ as $x\to 0$ and $\phi(x)\sim x^{-3/2}$ as $x\to\infty$ if 
$\chi(k=0)\not=0$. 
Note, that $\chi(k)\geq0$, hence $\phi(x)$ is a monotonic decreasing function.
One may keep in mind some particular form
of $\chi(x)$, 
say $\exp(-x^2L^2/2)$. Then, $\phi(x)=(1+x)^{-3/2}$. 

Now we derive a closed equation which describes the evolution of $w$ which is
the square root of the solution inverse width.  One can do that directly from
(\ref{i2}) substituting there $\sigma=w'(\tau)/2$ which is the consequence of
(\ref{ata}). One obtains an integral equation which is equivalent to some
third-order ordinary differential equations. The order can be then reduced by
one due to the conservation law (\ref{energ}). However, it is more instructive
to derive the same equation on $w$ in a different manner: since we are looking
for the extremum of the action (\ref{i1}) we can substitute there all the
fields as the functionals of $w$ and then make a variation with respect to $w$.
We have
\begin{eqnarray}&&
i{\cal I}=\frac{n}{2}\ln\left[
\frac{n}{e}\int_0^{\tau_0}
\frac{d\tau}{w^2(\tau)}
\phi\left(\frac{\tau}{Pe^2w(\tau)}\right)
\right]\nonumber\\&&-\frac{1}{4}\int_0^{\tau_0}
\frac{d\tau}{w(\tau)}\left(\frac{1}{2}w'-1\right)^2
\label{ac3}\end{eqnarray}
Varying with respect to $w$, we obtain
\begin{eqnarray}
w''-\frac{w'^2}{2w}+\frac{2}{w}-&&
\frac{8|y|^2}{w^2}\phi\left(
\frac{\tau}{Pe^2w}
\right)\nonumber\\&&-\frac{4|y|^2\tau}{Pe^2w^3}\phi'\left(
\frac{\tau}{Pe^2w}
\right)=0
\label{sec}\end{eqnarray}
Equation (\ref{sec}) can be rewritten as a Hamiltonian system
with the momentum $P=w'/w$ and the Hamiltonian
\begin{eqnarray}&&
H=\frac{P^2w}{2}+\frac{4|y|^2}{w^2}\phi\left(
\frac{\tau}{Pe^2w}
\right)-\frac{2}{w}.
\label{ham}\end{eqnarray}
Initial conditions for (\ref{sec}) are $w(0)=1$ and $w'(0)=-2(2n-1)$.  The
latter is readily derived from (\ref{i2},\ref{nlc}).  We should satisfy also a
final condition.  Indeed, (\ref{t},\ref{ac3}) require $w'\to 2$ as $\tau\to
\tau_0$, otherwise the integral contribution to the action is infinite.  Our
aim is to find such a value of $y$ that the solution of (\ref{sec}) satisfies
the relation (\ref{nlc}).  In other words, we can divide the task into two
parts. First, to find solution of equation (\ref{sec}) with arbitrary $y$ and
the given boundary conditions. Second, given $w$, to solve the algebraic
equation (\ref{nlc}) which determines $y$. Note that finite positive value of
$w'$ at $\tau\to \tau_0$ implies $\tau_0=\infty$.

Let us start the first part of our program, which is solving (\ref{sec}). Since
(\ref{ham}) explicitly depends on $\tau$, then $H$ is not conserved
and (\ref{sec}) can not be solved explicitly for an arbitrary
$\phi$.  In the limit $n\ll Pe^2$, it is possible nevertheless to describe the
solution with enough details to recover it's dependence on the parameters $n$
and $Pe$.  

The evolution of $w$ can be divided into three parts. During the initial stage,
when $\tau$ is close to zero, $w$ is of order unity. Therefore,
$Pe^2 w\ll \tau$ and we can substitute $\phi$ in (\ref{ham}) by its
asymptotic value $1$. Thus, during that stage $H$ is a constant which we denote
$H_0$. One finds $H_0=H(0)= 2(2n-1)^2-2+ 4|y|^2$. Since $n\gg1$ and, as we
shall see below $y\ll 1$, we have $H_0\approx 8n^2$. The equation for $w$ can
be then readily derived from (\ref{ham})
\begin{eqnarray}&&
w'=-\sqrt{4+2H_0w-8|y|^2/w}
\label{frst}\end{eqnarray}
Now let us consider the final stage. Since $w'\to 2$ as $\tau\to\infty$, we can
write $w\approx2\tau$. It gives $\tau/(Pe^2w)\approx1/(2Pe^2)\ll 1$. Thus, like
for the first stage, we can replace $\phi$ by $1$, and therefore the energy is 
a constant. It follows from (\ref{ham}) that during that stage $Hw\ll1$. 
Hence, $w$ satisfies the
following equation 
\begin{eqnarray}&&
w'=\sqrt{4-8|y|^2/w}
\label{thrd}\end{eqnarray}

Since $w'$ is negative during the first stage and positive during the third
one, then it has to turn into zero at some reflection time $\tau_*$. Around
that time, there should exists an intermediate part of evolution, which matches
the two above asymptotics.  We will see below, that this stage gives the main
contribution into (\ref{nlc}).  During that stage one has $\tau/(Pe^2w)\gtrsim
1$, so that $H$ is not conserved but decreases from $2n^2$ to $0$. It will be
important for us that $H$ is a decreasing monotonic function of $\tau$, which
becomes obvious after one differentiates (\ref{ham}).  

Let us make estimations of the parameters during that intermediate stage.  From
(\ref{frst}) it is easy to find that $w$ diminishes from $1$ to a substantially
smaller value during the time $\tau_*\simeq 1/n$. To have
$\tau/(Pe^2w)$ of order unity by the beginning of the second stage,
there should be $w\sim 1/(nPe^2)$.  Looking at (\ref{ham}), we observe,
that since $H$ is a decreasing function of $\tau$, the left-hand side of 
(\ref{ham}) is less than $H_0\approx 8n^2$. On the other hand, the term $2/w$
in the right-hand side can be estimated as $nPe^2$. If $n\ll Pe^2$, we can
disregard the left-hand side. We assume in addition that the duration of the
second stage is much less than $\tau_*$. Then, we can substitute $\tau$ by
$\tau_*$ in the argument of $\phi$ and obtain the
equation
\begin{eqnarray}&&
w'^2=4-\frac{8|y|^2}{w}\phi\left(
\frac{\tau_*}{Pe^2w}
\right)
\label{scnd}\end{eqnarray}
We do not take square root, since $w'$ changes sign during that
part of the evolution so that both branches of (\ref{scnd}) are pertinent. 
Actually, (\ref{scnd}) is valid for all $\tau>\tau_*$, since at $w\gg
\tau_*/Pe^2$ it turns into (\ref{thrd}). If our assumption about the duration
of the second stage is correct, the transition region is described correctly
too. On the other hand, if we add into (\ref{scnd}) the term $H_0 w$, which is
small in the transition region, the equation thus obtained will correctly
describe the evolution of $w$ at all times $\tau<\tau_*$.

Therefore, we found that under the assumption of a short transition region,
it is possible to reduce (\ref{sec}) to the equations
$w'=-\sqrt{\psi_1(w)}$ at $\tau<\tau_*$ and $w'=\sqrt{\psi_2(w)}$ at
$\tau>\tau_*$, where
\begin{eqnarray}&&
\psi_1=4+H_0w-\frac{8|y|^2}{w}\phi\left(
\frac{\tau_*}{Pe^2w}
\right),\nonumber\\&&
\psi_2=4-\frac{8|y|^2}{w}\phi\left(
\frac{\tau_*}{Pe^2w}
\right).\nonumber
\end{eqnarray}

Now we can continue with the second task, that is solving (\ref{nlc}). We
introduce also $w_*=w(\tau_*)$ at the moment of reflection.  Since at
$\tau=\tau_*$ the derivative of $w$ is zero, then we find from (\ref{scnd}) 
\begin{equation}
|y|^2=\frac{w_*}{2\phi\left(\frac{\tau_*}{Pe^2w_*}\right)}
\label{y2}\end{equation}
Now (\ref{nlc}) can be rewritten in the following form
\begin{eqnarray}&&
\frac{2n}{w_*}\phi\left(
\frac{\tau_*}{Pe^2w_*}
\right)=\int_{w_*}^1\frac{dw
}{w^2\sqrt{\psi_1(w)}}\phi\left(
\frac{\tau_*}{Pe^2w}
\right)\nonumber\\&&+
\int_{w_*}^\infty\frac{dw}{w^2\sqrt{\psi_2(w)}}\phi\left(
\frac{\tau_*}{Pe^2w}
\right)
\label{y}\end{eqnarray}

Estimating contributions into the integrals from the first and third time
intervals, one can find that alone, they are too small to satisfy (\ref{y}). On
the other hand, it is easy to see, that if the derivative of $\psi_{1,2}$ at
the point $w_*$ is not small enough, the second stage also does not contribute
into the integrals in (\ref{y}). The only way to have a solution is to make the
derivatives small. Then $\psi_{1,2}\approx \psi_{1,2}''(w_*)(w-w_*)^2/2$ in a
wide interval, and both integrals in (\ref{y}) logarithmically diverge, with
cut off on non-zero value of the first derivative. Since we can make
$\psi'(w_*)$ arbitrary small by a small change of $w_*$, the solution exists. 
Equating the derivative of $\psi$ to zero one finds, that $w_*$ is close to
$\alpha\tau_*/Pe^2$, where $\alpha$ is some number of the order unity, which
depends on the form of $\phi$ that is of the pumping $\chi$.  The deviation of
the first derivative from zero $\delta$ is determined by (\ref{y}):
\begin{eqnarray}&&
nw_*=\sqrt{\frac{2}{\psi''(w_*)}}\ln\frac{1}{\delta}\ .
\end{eqnarray}
One finds $\ln\delta^{-1}\propto n$. Smallness of $\delta$ justifyies our 
assumption
on a short intermediate stage.
Thus, we have shown that solution of the equation (\ref{nlc}) exists, and 
$|y^2|\sim 1/(n Pe^2)$ which corresponds to $\theta_x(0,0)\sim n^{3/2}Pe$.
This gives the main contribution into $\langle\theta_x^n\rangle\propto n^{3n/2}
Pe^n$ since the integral term in the action ${\cal I}$ is 
$\sim n$ on our instanton solution. Such moments correspond to the
following tail of the PDF of $\omega=\theta_x$ (note that it does not depend on
the strain amplitude $D$)
\begin{eqnarray}&& 
\ln \left[{\cal P}(\omega)\right]\propto -(\kappa\omega^2/P)^{1/3}.  
\label{tail}\end{eqnarray}

Let us describe the above instanton solution in more physical terms.  The
instanton corresponds to some optimal process giving the main contribution into
$\left\langle\theta_x^n\right\rangle$. It produces large gradient
which compensates for a small probability of such a process.  Looking at
(\ref{th1},\ref{tt}), we can distinguish the three stages of $\theta$
evolution. During the first one, starting at $t=-\infty$, the strain
$\sigma\approx D$ stretches small-scale initial perturbation up to the
width of order $L$
and the force prepares some profile of $\theta$ which has the amplitude of
order $\sqrt{P/D}$. The second stage starts when $w$ is
close to $w_*$. Then $\sigma\ll D$, we can disregard both the advective and
diffusive terms.  Therefore, the width of $\theta$ does not 
change during that stage while the
amplitude grows due to the force.  Of all realizations of the force the
constant one is preferred, since it gives the fastest growth. Then, $\theta$
increases as $\phi\, t$. The weight of such a process is $\exp(-\phi^2t-t/2)$. 
The second term in the exponent is the probability to have small $\sigma$
during the time $t$. Then we can find that $\phi\sim1$ and $t\propto n$ (note 
that the second stage is long in terms of $t$ yet it was short in
terms of $\tau$). By the
end of this stage $\theta_x\propto n$. And finally, during the last stage we
can disregard the force. The profile having the amplitude
$n$ and width $L$ by the beginning of the stage is compressed by the large
negative $\sigma$ which can be estimated as $\sigma\sim -Dn$.  
The duration of that
stage (and the final width) is determined by diffusion: $\sigma
x\partial_x\theta\sim\kappa\partial_x^2\theta$ at the end. Then, the width of
$\theta(x)$  is $\sqrt{\kappa/Dn}$ while the amplitude is $n\sqrt{P/D}$, 
therefore the final answer for $\langle\theta_x^n\rangle$ is $\propto
n^{3n/2}(P/\kappa)^{n/2}$, which corresponds to (\ref{tail}). To summarize, the
optimal fluctuation that gives the main contribution into
$\langle\theta_x^n\rangle$ starts from an infinitesimal fluctuation
which is initially stretched, then it has a long stage of suppressed advection 
when
the amplitude of $\theta(x,t)$ grows yet it's spatial scale does not decrease,
and then it is contracted fast.

Let us stress that the instanton describes a very special configuration of the
fields. For any other solution which does not satisfy correct self-consistency
condition, there is no such a long intermediate stage of growth. If the
parameters were not fitted to guarantee the existence of this long stage,
either large $\sigma$ would bring a singularity in the solution or $\theta_x$
would not be large enough at $t=0$.

Note that the scalar itself has an exponential PDF tail, the fact that the
gradient PDF is less steep was correctly attributed in \cite{97CKVa} to the
fluctuations of the diffusion scale. One may explain $2/3$ stretched exponent
in the following simple way: Due to diffusion, local gradient can be thought of
as a product of a scalar fluctuation and inverse diffusion scale. The former
has an exponential PDF tail \cite{97CKVa} while the latter is proportional to
the local stretching rate which is Gaussian.  Therefore:
$\langle\omega^n\rangle\sim \langle\theta^2\rangle^{n/2}n^nr_d^{n}n^{n/2}$
which corresponds to (\ref{tail}). Also, instanton formalism provides for an
instructive insight into the relation between the scalar and it's gradient on
the optimal fluctuation: comparing the second and the third terms in (\ref{e1})
one gets for the current dissipative scale $r_d\propto \sqrt{\kappa/\theta D}$,
substituting that into $\theta_x\simeq\theta/r_d$ we obtain $\theta_x\propto
\theta^{3/2}$ so that exponential tail for $\theta$ has to correspond to $2/3$
stretched exponential for the gradient.

To conclude, the gradient's PDF is given by (\ref{ad}) for
$\kappa\omega^2/P\ll1$ and by (\ref{tail}) for $1\ll\kappa\omega^2/P\ll Pe^3$
which agree with the results found by a time-separation formalism
\cite{97CKVa}. Speaking on generalizations, it is likely that one-dimensional
instanton described here may be relevant for a multidimensional case, both
compressible and incompressible, due to universality of a locally flat
ramp-and-cliff structure discussed in \cite{91Sre,HS94,P94}. Note that the
stretched exponential tail is what one expects for steady gradient's
distribution (which is possible only when diffusion is present) \cite{94SS}
contrary to unsteady log-normal distribution which takes place without
diffusion \cite{74Kra-a}.  

We are indebted to M. Vergassola for numerous useful
explanations and helpful remarks.
We are grateful to V. Lebedev for a valuable remark made upon reading 
the first draft of the 
paper. We thank M. Chertkov and I. Kolokolov for useful
discussions. We thank U. Frisch and L. Biferale for organizing an
inspiring workshop at Nice where this work has been started. G.F. 
thanks K. Gawedzki for a kind hospitality and a very stimulating scientific
atmosphere at Bures-sur-Yvette, where this work was continued.  The work was
supported  by the Israel Science Foundation
and the Minerva Center for Nonlinear
Physics at the Weizmann Institute.

\end{multicols}

\end{document}